\documentclass[aps,prd,nofootinbib]{revtex4}
\usepackage{latexsym}
\usepackage{graphicx}
\usepackage{amsmath,amsfonts}
\usepackage{amsbsy}
\usepackage{mathrsfs}
\usepackage{color}
\usepackage{psfrag}
\usepackage{enumerate}
\usepackage{amsmath,amssymb,calc,amsfonts}
\usepackage{latexsym}
\DeclareFontFamily{U}{rsfs}{}         
\DeclareFontShape{U}{rsfs}{m}{n}{<5> rsfs5 <6><7> rsfs7          %
  <8><9><10><10.95><12><14.4><17.28><20.74><24.88> rsfs10}{}     %
\DeclareMathAlphabet{\mathfs}{U}{rsfs}{m}{n}                     %
\definecolor{indiagreen}{rgb}{0.07, 0.53, 0.03}
\def\beq{\begin{eqnarray}}
\def\eeq{\end{eqnarray}}
\def\l{\ell}
\def\mb#1{\mathbb{#1}}

\def\nn{\nonumber\\}

\def\mb#1{\mathbb{#1}}

\def\={\stackrel{\Delta}{=}}

\def\mbf#1{\mbox{\boldmath${#1}$}}

\def\half{{\textstyle{\frac{1}{2}}}}

\begin{document}

\title{Exponential corrections to black hole entropy}

\author{Ayan Chatterjee}\email{ayan.theory@gmail.com}
\affiliation{Department of Physics and Astronomical Science, Central University 
of Himachal Pradesh, Dharamshala 176215, India.}
\author{Amit Ghosh}\email{amit.ghosh@saha.ac.in}
\affiliation{Theory Division, Saha Institute of Nuclear Physics, 1/AF Bidhan Nagar, Kolkata 700064, India,\\
 Homi Bhabha National Institute, Anushakti Nagar, Mumbai 400094, India.}
\begin{abstract}
Using the quasilocal properties alone we show that the area spectrum of a black hole horizon must be 
discrete, independent of any specific quantum theory of gravity. The area spectrum is found to be half-integer 
spaced with values $8\pi \gamma \ell_{p}^{2}j$ where $j\in \mathbb{N}/2$. We argue that if microstate counting is carried out for 
quantum states 
residing on the horizon only, correction of $\exp(-\mathcal{A}/4\ell_{p}^{2})$
over the Bekenstein-Hawking area law must arise in black hole entropy. 
\end{abstract}
\maketitle
According to our present understanding black hole horizons are identical to thermodynamic systems. 
The classical dynamics of black hole horizons encode thermal behavior. Isolated black hole horizons in equilibrium indeed
have constant surface gravity ($\kappa$) and their classical evolution from one equilibrium state to another change 
the parameters like mass ($M$), angular momentum (${\mathcal{J}}$), etc. in such a way that a relation identical to 
the first law of thermodynamics is obeyed provided horizons are assigned a temperature $T=\hbar\kappa/2\pi$ and
the horizon area $\mathcal{A}$ is equated to the 
thermodynamic entropy $S=(\mathcal{A}/4\ell_{p}^{2})$ where $\ell_{p}$ is the Planck length. The law 
of ever increasing classical area $\cal A$ enforces the analogy further \cite{Bardeen:1973gs,Hawking:1974sw,Bekenstein:1973ur}.

The study of the microscopic origin of entropy is a
major thrust area of quantum black hole physics. It is generally expected that any quantum theory of black hole must 
furnish an explanation of the Bekenstein-Hawking area law for entropy. Microstate counting in string theory as well as 
loop quantum gravity (LQG) not only 
yield the Bekenstein-Hawking area law but also produce corrections to it (as an expansion of $\ell_P^2/\mathcal{A}$) including a logarithmic term
\cite{Strominger:1996sh, Dijkgraaf:1996it, Dabholkar:2008zy,
Mandal:2010cj, Ashtekar:1997yu,Kaul:2000kf,Meissner:2004ju,Domagala:2004jt, Ghosh:2004wq}.
These corrections appear for horizons having areas large compared to $\ell_{p}^{2}$.
It is now accepted that black hole entropy should have the following form:
\begin{equation}
S=\frac{\mathcal{A}}{4\ell_{P}^{2}}+\alpha\ln\frac{\mathcal{A}}{4\ell_{P}^{2}}+\beta \frac{4\ell_{P}^{2}}{\mathcal{A}}+\cdots 
+\exp\left(-\delta\,\frac{\mathcal{A}}{4\ell_{P}^{2}}\right)+ \cdots ,
\end{equation}
where $\alpha, \beta, \delta$, etc. are universal constants.
For small horizons having areas $O(\ell_{p}^{2})$ (whose understanding require a full theory of quantum gravity) the log and subsequent correction terms involving $(\ell_P^2/\cal A)$ and its higher orders may be either absent or modified. Indeed, it has been stressed in \cite{Ghosh:2012jf} that even for large areas logarithmic corrections do not arise if the microstate counting is made in a certain manner. 
Clearly even if these terms are inescapable, they must be negligible in the small area limit. However the exponential term is interesting: although negligible for large areas, it may become an important correction if the area is small. 
The exponential correction has not been studied in the literature at length although some interesting computations
in string theory exhibit such terms \cite{Dabholkar:2014ema}. In this paper 
we shall derive black hole entropy and the exponential correction
using \emph{only} the horizon geometry and without appealing either to string theory or LQG. In the process we shall identify local horizon microstates and also derive an area spectrum. Our results indicate that exponential corrections in black hole entropy may arise in any quantum theory of gravity.

An important feature of our approach which will play a crucial role in our derivation lies in the quantum representation of black hole horizons. 
Note that quantum descriptions of black holes in string theory or LQG make use of the entire spacetime. 
The use of bulk is explicit in LQG and implicit in string theory (the quantised brane configurations are expected to reproduce 
the entire spacetime not the horizon only). In some sense these characterizations require the horizon to be quantum mechanically entangled with 
the bulk although classically it remains isolated. Instead we develop a picture of a horizon 
which remains classically isolated and does not communicate with the bulk even quantum mechanically. 
The picture is somewhat like an individual atom whose quantum theory does not force it to interact with 
the rest of the universe and its quantum states are not necessarily entangled with its surroundings. This leads
to a truly \emph{isolated quantum black hole} subjected to microstate counting. 


In four-dimensional spacetime $\mathcal{M}$ a black hole horizon in equilibrium ($\Delta$) is best described by a
Weak Isolated Horizon (WIH) \cite{Ashtekar:2000sz, Ashtekar:2004cn,Perez:2017cmj}.
$\Delta$ is a null hypersurface in $\mathcal{M}$ such that its generator $\ell^{a}$ (belonging
to an equivalence class $[\xi \ell^{a}]$, $\xi$ being a function on the horizon) is a null vector field 
which is shear-free, expansion-free and Killing on the horizon. 
The acceleration of $\ell^{a}$ obtained from $\ell^{a}\nabla_{a}\ell^{b}=\kappa_{(\ell)}\ell^{b}$ is called the surface gravity $\kappa_{(\ell)}$. Since $\Delta$ is defined without reference to asymptotic infinity,
$\kappa_{(\ell)}$ is local. By appropriate choices of the function $\xi$ the horizon $\Delta$ admits  
all possible values of surface gravity including zero for extremal horizons. 
We assume that $\Delta$ is topologically $\mathbb{S}^{2}\times \mathbb{R}$.
The null vectors $(\ell^{a}, n^{a}, m^{a}, \bar{m}^{a})$ will be used as the spacetime basis 
(the non-zero dot products being $\ell\cdot n=-1, m\cdot \bar{m}=1$). In this basis the spacetime metric 
is given by $g_{ab}=-2\ell_{(a}n_{b)}+2m_{(a}\bar{m}_{b)}$. 

In this paper, we will be interested in internal Lorentz symmetries
and hence use the first order tetrad-connection variables because
metric variables cannot disentangle diffeomorphisms from
Lorentz transformations. The tetrad variable $e_{a}^{I}$ maps the spacetime vector fields 
to internal flat Minkowski space vectors $\ell^{I}=e^{I}_{a}\,\ell^{a}$ ($a,b, \dots$ are spacetime indices while 
$I,J, \dots$ are for internal flat spacetime). The connection one-form 
$(A_{aI}{}^{J})$ is defined by,
$\nabla_{a}\lambda^{I}=\partial_{a}\lambda^{I}+A_{a}{}^{I}{}_{J}\lambda^{J}$ where $\lambda_{I}$ is
an internal vector and $\partial_{a}$ is the 
internal flat connection. The horizon $\Delta$ will also be assumed to have a fixed set 
of internal tetrad basis $(\l^{I}, n^{I}, m^{I}, \bar{m}^{I})$ annihilated by
the internal flat connection. In the bulk the spacetime $\mathcal{M}$ allows all possible Lorentz transformations ($\Lambda^{I}{}_{J}$) of the
tetrads $e_{a}^{I}$. However on $\Delta$ two criteria must be satisfied: first the vector field $\ell^{a}=e^{a}_{I}\ell^{I}$
should belong to the equivalence class $[\xi\ell^{a}]$ and second only those $SL(2,C)$ transformations are allowed 
which preserve the boundary conditions on $\Delta$. These transformations constitute the `symmetries' of the $\Delta$
since they either preserve the Newman-Penrose coefficients on the horizon or transform them homogeneously. 
The generators of these symmetries are \cite{Chatterjee_ghosh_basu}
\begin{eqnarray} 
&B_{IJ}=-2\l_{[I}n_{J]}, ~~P_{IJ}=2m_{[I}\l_{J]}+2\bar m_{[I}\l_{J]}\nn
&R_{IJ}=2im_{[I}\bar m_{J]}, ~~Q_{IJ}=2im_{[I}\l_{J]}-2i\bar m_{[I}\l_{J]},\label{lbq}
\end{eqnarray}
where $R$ generates Euclidean rotations
in the ($m$-$\bar{m}$) plane, $P$ generates rotation in $(\ell$-${m})$ plane,
$Q$ generates rotation in $(\ell$-$\bar{m})$ plane and $B$ generates scaling transformations of $\ell$ and $n$. 
These generators obey the Lie algebra of $ISO(2)\ltimes\mb R$ where the symbol $\ltimes$ stands for the semidirect
product. $R,P$ and $Q$ generate $ISO(2)$ on $\mathbb{S}^2$ while $B$ generates $\mb R$
\begin{eqnarray}\label{vfalgebra}
&&[R,B]=0,\quad [R,P]=Q,\quad [R,Q]=-P,\nonumber\\
&&[B,P]=P,\quad [B,Q]=Q,\quad [P,Q]=0,\label{iso2}
\end{eqnarray} 
where $[R,B]_{IJ}=R_{IK}B^K{}_J-B_{IK}R^K{}_J$. This is not surprising
since $ISO(2)$ is the little group of the Lorentz group that keeps the
horizon generator invariant. 

We consider a spacetime region bounded by $\Delta$, two Cauchy surfaces $M_\pm$ respectively denoting the future and past boundaries and the asymptotic boundary.
We assume suitable fall-off conditions on the fields at asymptotic boundary  
for a well defined action principle. In this region of spacetime the transformations
generated by $ISO(2)\ltimes\mb R$ map fields to their equivalent configurations and hence are pure gauges. However at the boundary $\Delta$, these symmetries may acquire the status of a global transformation and give rise to physical charges. It is well known that in presence of boundaries
local symmetries may lead to observable charges and examples like edge states of gauge theories arise in this way. Another familiar example is Chern-Simons theory on a three-manifold with boundary, 
say a disc $\mathbb D\times \mathbb{R}$, with $\mathbb{R}$
playing the role of time. In this case gauge transformations take field configurations
in the bulk to their gauge equivalent ones but on the boundary become global symmetries \cite{witten}. 
In gravity too the gauge motions due to diffeomorphisms relate gauge equivalent geometries in the bulk but 
they become genuine symmetries on the boundary giving rise to observable charges \cite{Szabados}. Similarly for the Lorentz transformation belonging to $ISO(2)\ltimes\mb R$, the Hamiltonian generator or the phase space charge is expected to become a physical charge on the horizon.

To determine the Hamiltonian charges for internal Lorentz symmetries we 
use the Holst action in ($e_{a}^{I},\, A_{aI}{}^{J}$) variables. 
It is classically equivalent to the Einstein-Hilbert action in second order metric variables.
The Holst action is given by the following Lagrangian (the factor $16\pi G\gamma$ is a constant) \cite{Holst:1995pc,Chatterjee:2008if}:
\begin{equation}\label{lagrangian1}
-16\pi G\gamma\,L= \gamma \Sigma_{IJ}\wedge F^{IJ}-e_{I}\wedge e_{J}\wedge F^{IJ},
\end{equation}
where $\Sigma^{IJ}=\half\,\epsilon^{IJ}{}_{KL}e^K\wedge e^L$, $A_{IJ}$ is a Lorentz $SO(3,1)$ connection 
and $F_{IJ}$ is a curvature two-form corresponding to the connection given by
$F_{IJ}=dA_{IJ}+A_{IK}\wedge A^{K}~_{J}$. It is useful to add the boundary terms
$[d(e_{I}\wedge e_{J}\wedge A^{IJ})-\gamma\,d(\Sigma_{IJ}\wedge A^{IJ}) ]$ to the Lagrangian
to make calculations simpler \cite{Chatterjee:2008if}. 
The covariant phase space for this Lagrangian contains all the solutions of the (\ref{lagrangian1}) which allow $\Delta$
as inner boundary. Well-known black hole solutions including the Schwarzschild and Kerr
belong to this space of solutions. The symplectic structure on this space of solution
has contributions from the spacetime bulk and the boundary:
\begin{eqnarray}\label{Palatini_1}
(16\pi G\gamma)\, \Omega(\delta_{1}, \delta_{2}
)=\int_{M}\,\delta_{[1}(e^{I}\wedge
e^{J})~\wedge\delta_{2]}A^{(H)}_{IJ} 
+\int_{S_{\Delta}}\,
\delta_{[1}{}^2\mbf{\epsilon}~\,\delta_{2]}\{\mu_{(m)}+ \gamma\psi_{(\l)}\} ,
\end{eqnarray}
where $M$ is a partial Cauchy slice that intersects the horizon $\Delta$ at the sphere $S_{\Delta}$ 
and $\delta_1,\delta_2$ are vector fields on the phase space. The quantity $A^{(H)}_{IJ}=(1/2)[\,A_{IJ}-(1/2)\,\epsilon_{IJKL}\,A^{KL}]$
and $\psi_{(\ell)}$ and $\mu_{(m)}$ are phase space functions \cite{Chatterjee:2008if}.
The quantity ${}^2\mbf{\epsilon}$ is the area two-form on the spherical cross sections $S_{\Delta}$ of the horizon. The fields 
$\psi_{(\ell)}$ and $\mu_{(m)}$ are assumed to satisfy the boundary condition that $\psi_{(\ell)}=0$ and $\mu_{(m)}=0$ at 
some initial cross section of the horizon. We shall also use the result \cite{Lewandowski:2018rzc}
that for certain class of spacetimes (Bardeen-Horowitz class) which are solutions of Einstein's equations with 
possibly nonzero cosmological constant may be foliated by expansion-free, twist-free null surfaces generated by null-vector 
field $\ell^{a}$. These surfaces are transverse to a fiducial extremal null horizon placed at
$v=-\infty$ in the advanced Eddington-Finkelstein coordinates. The Cauchy surface $M$ cuts through these foliation
surfaces. Using this result we obtain the tetrad products in the full spacetime in the basis of $ISO(2)\ltimes\mb R$ 
\begin{eqnarray}
e^{I}{}_{a}\wedge e^{J}{}_{b} &=&-2~n_{a}\wedge m_{b}~\l^{[I}\bar m^{J]} -2~n_{a}\wedge \bar m_{b}~\l^{[I} m^{J]} +2i~m^{[I}\bar m^{J]}~{}^2\mbf{\epsilon}_{ab}\nn
\Sigma_{ab}{}^{IJ}&=& 2\l^{[I}n^{J]}~ {}^2\mbf{\epsilon}_{ab} +2n_{a}\wedge(im_{b} \l^{[I}\bar m^{J]} - i\bar m_{b}\l^{[I} m^{J]}).
\end{eqnarray}
The connection one-form is given in the basis of the algebra of $ISO(2)\ltimes\mb R$ and has the following form \cite{Chatterjee:2008if}
\begin{equation}
 A_{IJ}=-2~\omega^{(\ell)}~ \ell_{[I}n_{J]}+
2~U^{(l,m)}~\ell_{[I}\bar m_{J]}+ 2~{\bar U}^{(l,m)}~\ell_{[I}m_{J]}+ 2~V^{(m)}~m_{[I}\bar m_{J]}.
\end{equation}
To evaluate the symplectic structure, note that the variations of 
the tetrads and the connection due to infinitesimal Lorentz 
transformations $\Lambda^{I}{}_{J}=(\delta^{I}{}_{J}+\,\epsilon^{I}{}_{J})$
are given by: 
\begin{equation}
\delta_\epsilon e^I=\epsilon^{I}{}_{J}\,e^{J}\, ;\,~~~~
\delta_\epsilon A^{IJ}=d\epsilon^{IJ} +A^{IK}\epsilon_{K}{}^{J}+A^{JK}\,\epsilon^{I}{}_{K}\label{delta_a}.
\end{equation}
%

%

Using \eqref{delta_a} in the $\gamma$-independent (the Palatini) part of symplectic structure of \eqref{Palatini_1} leads to:
\begin{eqnarray}\label{omega_variation1}
\Omega_{B}(\delta_\epsilon,\delta)&=-&\frac{1}{8\pi G}\int_{M}(\epsilon^K~_I\Sigma_{JK}\,\wedge\delta A^{IJ}
-\delta\Sigma_{IJ}\wedge A^{IK}\epsilon_{K}{}^{J})-\delta\Sigma_{IJ}\wedge d\epsilon^{IJ},
\end{eqnarray}
where the subscript $B$ denotes the bulk part of the symplectic structure and the boundary part vanishes. 
The second term in \eqref{omega_variation1} may be rewritten as 
$\delta\Sigma_{IJ}\wedge d\epsilon^{IJ}=d(\delta\Sigma_{IJ}\,\epsilon^{IJ})+\delta(A_I{}^K\wedge\Sigma_{KJ}
+A_J{}^K\wedge\Sigma_{IK})\epsilon^{IJ}$.
Using these expressions in the symplectic structure \eqref{omega_variation1}, 
we note that the terms with $\delta \Sigma_{IJ}$ cancel each other while those with $\delta A_{IJ}$ cancel 
for the Lorentz transformations which belong to the symmetry group on a WIH. After some
algebra we obtain the following quantity on the cross-sections ($S_{\Delta}$) of the horizon:
\beq\label{bulk_term_charge_1}
\Omega_{B}(\delta_\epsilon,\delta)&=&\frac{1}{16\pi G}\int_{S_{\Delta}}\delta\Sigma_{IJ}\,\epsilon^{IJ}.
\eeq
Similarly, for the $\gamma$-dependent symplectic structure also a similar 
expression may be obtained.
%
%
The bulk contribution of the full Holst action to the symplectic structure is then reduces to:
\begin{eqnarray}\label{bulk_Holst_1}
\Omega_{B}(\delta_\epsilon,\delta)=- \frac{1}{16\pi G\gamma }\int_{S_{\Delta}}\delta\, (e_{I}\wedge e_{J}-\gamma \Sigma_{IJ})\wedge \epsilon^{IJ}.
\end{eqnarray}
For $\epsilon_{IJ}=R_{IJ}=2im_{[I}\bar m_{J]}$ the symplectic structure in \eqref{bulk_Holst_1} 
gives the Hamiltonian generating internal rotation in the phase space. Since there is only one rotation on 
the horizon, we shall denote it by $-J$ and the only contribution comes from the $\gamma$-dependent part of the symplectic structure:
\begin{eqnarray}
\Omega_{B}(\delta_R,\delta)=- \frac{1}{8\pi G\gamma }\int_{S_{\Delta}}\delta\,~ {}^{2}\mbf{\epsilon}
=- \delta\left(\frac{\mathcal{A}}{8\pi G\gamma }\right)\equiv\delta(-J).
\end{eqnarray}
So $(\mathcal{A}/8\pi G\gamma)$ is the generator of rotation in the phase space of isolated horizons.
For $\epsilon_{IJ}=B_{IJ}=-2\ell_{[I}n_{J]}$ we denote the charge by $K$ as it is a boost on the horizon and the only contribution comes from the $\gamma$-independent part of the symplectic structure \eqref{bulk_Holst_1}:
\begin{eqnarray}
\Omega_{B}(\delta_B,\delta)=\frac{1}{8\pi G }\int_{S_{\Delta}}\delta\,~ {}^{2}\mbf{\epsilon}
=\delta\left(\frac{\mathcal{A}}{8\pi G}\right)\equiv\delta(K).
\end{eqnarray}
Again $(\mathcal{A}/8\pi G)$ is the generator of boosts in the phase space of isolated horizons, generalising 
\cite{Carlip:1993sa,Massar:1999wg,Wall:2010cj,
Chatterjee:2015lwa}. 
One may also show that the Hamiltonian charges of the remaining two 
generators $P_{IJ}$ and $Q_{IJ}$ vanish on the horizon.  It also follows from this symplectic structure that the
the algebra of the Hamiltonian charges is identical to algebra of the spacetime vector fields. 
Thus we have derived two results of immense importance: First, the relation
$K=\gamma J$ which has important implications in quantum gravity and is usually referred to as the linear simplicity constraint
\cite{Rovelli:2013osa}. Second, the horizon area is linked with the internal angular momentum through
the relation $\mathcal{A}=8\pi G \gamma J$. 
In the following, we show that the quantum states residing on the horizon belong to a finite dimensional representation of 
the Lie algebra of $ISO(2)$. These states are also the eigenstates of $J$ and are labeled by integers or half-integers and consequently the $\mathcal{A}$-$J$ relation implies that the spectrum of $\mathcal{A}$ is naturally discrete.

Let us now identify the quantum states on the horizon cross-section. We note that
the algebra of vector fields is faithfully mapped to algebra of charges on the horizon.
If the generators corresponding to
$P_{IJ}$ and $Q_{IJ}$ are denoted by $Q$ and $P$ respectively (to make the algebra similar to the algebra (\ref{vfalgebra})) the quantum algebra is
\begin{equation}
[\,J,\,P\,]=i\hbar Q, ~~ [\,J,\,Q\,]=-i\hbar P,~~ [\,P,\,Q\,]=0.
\end{equation}
The operator $\mathcal{P}^{2}\equiv P^{2}+Q^{2}$ commutes with the algebra. If the eigenvalues of $\mathcal{P}^{2}$ and $J$ 
are $p^{2}$ and $j$ respectively then the states are labeled by $|p^{2},j\rangle$. 
Linear combinations of $P,Q$ form the shift operators:
$P_{\pm}=P\pm iQ$. A simple algebra shows that $P_{\pm}$ are
the raising and lowering operators respectively. More precisely
$P_{\pm}|p^2,j\rangle=\hbar|p^2,j\pm1\rangle$.
In case of WIH the generators corresponding to $P_{IJ}$ and $Q_{IJ}$ must vanish
and hence for solutions belonging to the WIH phase space both $P_{+}$ and $P_{-}$ vanish. In other words the label $j$ of 
the states are not raised or lowered and the operators $P_\pm$ act as constraints on the physical states $P_{\pm}|p^2,j\rangle=0$. So 
the physical states of the horizon must have $p^{2}=0$ and labeled by $j$ \emph{alone}. This is consistent with the homogeneous action of 
rotation operator on $P$ and $Q$, since these rotated vector operators shall continue to have vanishing eigenvalues. 
Hence, the irreducible representations for this case are one-dimensional and states are labeled by integer or half-integer $j$
\cite{Weinberg}. Note that these states are independent of quantum states residing in the bulk.

The analysis shows that on a WIH phase space the eigenstates of $J$ may be used 
to determine the spectrum of the area operator $\mathcal{A}|j\rangle=8\pi G \gamma J\,|j\rangle=8\pi G\hbar \gamma  j\, |j\rangle$. 
The area eigenvalues, also denoted by $\mathcal{A}$, are then $8\pi G \gamma\hbar j$. This is similar to the result of \cite{Ashtekar:2004eh}.
In the present scenario the quantisation arises naturally from to geometry of the WIH.
Note that on a WIH the operators $P_\pm$ do not change $j$. Since $j$ gives the total area,
this implies that the operators and states defined here naturally incorporate the fact that the area of WIH should not
change.

For the microstate counting we first note that large area $\cal A$ corresponds to large $j$. Since a large $j$-representation can be built
from a large number of smaller $j$-representations, we assume that a large area is a sum of smaller areas. This gives 
the microscopic germs of the surface $S_{\Delta}$ as large number of tiles much like the tessellation on the
surface of a soccer ball. We further assume that the $j$-labels of the tiles are independent of each other, 
that is no further constraint is imposed on their sum. Although the tessellation is motivated by the representation theory, for now we do not 
have a good argument to support the assumption of independence of $j$s used to label the tiles. 
These assumptions are however testable if we quantise the WIH in a full quantum theory of gravity.
 Often a quantum theory also involve further assumptions and for now our assumptions may be regarded as simplest. Since
a quantum state of the full classical area $S_{\Delta}$ is labeled by an integer or half-integer $|j\rangle$, this implies that 
the area of each tile should also be labeled by integers or half-integers. The macrostate $|j\rangle$ is given by a 
tensor product $|j\rangle=\otimes_i|j_{i}\rangle$ where $i$ labels the tiles. 
The eigenvalue of the area operator is given by $\mathcal{A}=\oplus_i\mathcal{A}_{i}$ where each tile with label $j_{i}$
contributes an area ${\cal A}_i=8\pi \gamma \ell_{p}^{2}j_{i}$. Thus $j=\sum_{i} j_{i}$. This equation is the
basis for calculating the black hole entropy which is obtained by determining the number of independent ways 
the configurations $\{j_i\}$ can be chosen such that for a fixed $j$ the condition $j=\sum_ij_i$ is satisfied.
The choice of independent tiling is however subject to diffeomorphism constraints. Using arguments 
similar to LQG \cite{Ashtekar:2004eh} we may fix the diffeomorphism constraints by coloring 
the tiles. However this process of fixing the diffeomorphism gauge makes the tiles distinguishable. 
Suppose in the partition of $j=N/2$ the number $n_i=2j_i$ is shared by $s_i$ tiles. 
Then the $\sum_is_in_i=N$ and $\sum_is_i$ is the total number of tiles in the tessellation. So 
the total number of independent configurations is given by
\begin{equation}
\Omega=\frac{(\sum_is_i)!}{\prod_is_i!}.
\end{equation}
Varying $\log\Omega$ subject to the constraint $\delta\sum_is_in_i=0$ yields the most 
likely configuration $s_i=(\sum_is_i)\exp(-\lambda n_i)$ where the variation parameter $\lambda$ is to be 
determined from the constraint $\sum_i\exp(-\lambda n_i)=1$ where $n_i=1,...,N$. This
gives $\lambda=\ln 2-2^{-N}+o(2^{-2N})$ for large $N$ and entropy $S=\lambda N$. Substituting $N$ we get
%
%
\begin{equation}\label{entropy}
S=\frac{\mathcal{A}\ln 2}{8\pi\gamma\ell_{p}^{2}}+e^{-{\mathcal A}\ln 2/8\pi\gamma\ell_p^2}.
\end{equation}
Thus for the choice $\gamma=\ln(2)/2\pi$, the leading order Bekenstein-Hawking result is reproduced, 
but also an exponentially suppressed correction to the {\em classical} result is obtained. 
This is an unexpected result since the present {\em it from bit} formulation of horizon gives logarithmic corrections.
The exponential suppression has been shown to arise in 
some nonperturbative string computations \cite{Dabholkar:2014ema} but has not been found in LQG calculations. 
Note that the entropy calculation uses large value for $s_i$. However keeping in mind that Stirling's approximation 
holds well even for small numbers (for $n=2$ Stirling's approximation gives 1.91 and the difference is an order of magnitude smaller 
than $\ln 2$) this correction is expected to survive for small areas $O(\ell_{p}^{2})$ as well and fail only in the sub-Planckian regime.   

In summary we have reported two major results in this paper. First the classical boundary conditions of a WIH and 
symplectic structure of Einstein's theory together imply that the classical area of horizon is the
Hamiltonian charge or generator of internal rotation. The relation $J=\mathcal{A}/8\pi G\gamma$ is reminiscent of 
the well-known area quantisation in LQG where the classical horizon area is quantised by representations of 
the internal angular momentum operator $\sqrt{J^2}$. However we show that such a relation arises directly at 
the level of classical phase space of WIH. It is a new and unexpected result. It connects Einstein's theory of gravity, its internal 
rotational symmetries and classical black hole horizons in an intriguing way and relates
the classical area of a WIH to representation of the internal angular momentum operator and thus shows how quantisation of area occur. 
Although the area-spectrum is in variance with the LQG literature \cite{Ashtekar:2004eh} it agrees with 
one of the regularised version proposed in \cite{Alekseev} and also with \cite{dreyer} from quasinormal modes. 
Second by choosing an appropriate representation the quantum states of a WIH and counting
the most natural microstates of this representation corresponding to a given classical area
correctly reproduces the semi-classical result of entropy and predicts a 
new form of quantum correction. These corrections do not involve any logarithmic term as in other counting schemes
but falls-off exponentially from the semi-classical value. This is also a new result and is expected to hold up to the Planckian 
regime of area $O(\ell_{p}^{2})$. To probe into sub-Planckian regime one has to 
do an exact counting of microstates without employing Stirling's approximation.
We reiterate that so far symmetry has been our sole guiding principle and the tessellated description is 
only a plausible model of microstates on the horizon. In a full theory of quantum gravity these notions can be 
tested but one needs to make further assumptions about the quantum theory itself such as the Hilbert space, operators, etc. and 
also about the classical limit in which the WIH phase space emerges. The black hole horizon used in LQG is very similar in spirit to
this model but there are differences in details such as the bulk-boundary constraint which plays a major role in quantizing a WIH. 
Our microscopic model should be viewed as the simplest one which relies on the geometric properties of the horizon alone and accounts for
the black hole entropy. In the future, we wish to carry out a detailed investigation of 
the phase space and Hamiltonian charges in a quantum theory of gravity.

%

\subsection*{Acknowledgements}
The authors are also supported by the Department of Atomic Energy, BRNS project grant 58/14/25/2019-BRNS.
AC is also supported by SERB-DST through their MATRICS project grant MTR/2019/000916. AC also thanks IUCAA for 
a visit through its Visiting Associate Programme. 

%

\end{document}